\RequirePackage{ifpdf}
\documentclass{PoS}

\usepackage{cancel}
\usepackage{graphicx}
\usepackage{amsmath}
\usepackage{amssymb}
\usepackage[mathscr]{eucal}
\usepackage{fancyvrb} 
\usepackage[hyphens]{url} 
\usepackage[pdftex]{color}     

\def \ar      {\texttt{AMBRE}{}}
\def \pltest      {\texttt{PlanarityTest}{}}
\def \mbn      {\texttt{MBnumerics.m}{}}

\def \mbsums  {\texttt{MBsums~}{}}
\def \la {\texttt{LA}{}}
\def \ga {\texttt{GA}{}}
\def \mb {\texttt{MB}{}}

\def \mbm  {\texttt{MB.m}{}}

\title{%
{\flushright{ 
\small \texttt{DESY 16-120}
\\
\small \texttt{KW 16-01}
\\[8mm]
}}
Numerical integration of massive two-loop \\Mellin-Barnes integrals in Minkowskian regions}

\ShortTitle{Numerical integration of massive two-loop Mellin-Barnes integrals in Minkowskian regions}

\author{Ievgen Dubovyk$^{a,b}$, 
\speaker{Janusz Gluza}$^{,c}$,
Tord Riemann$^{a,c,d}$, 
Johann Usovitsch$^a$
\\
\\
\llap{$^a$}  Deutsches Elektronen-Synchrotron, DESY, Platanenallee 6, D-15738 Zeuthen, Germany 
\\
\llap{$^{b}$} II. Institut f{\"u}r Theoretische Physik, Universit{\"a}t Hamburg,
22761 Hamburg,  Germany 
\\
\llap{$^c$} Institute of Physics, University of Silesia, Uniwersytecka 4, PL-40007 Katowice, Poland 
\\
\llap{$^d$} 15711 K{\"o}nigs Wusterhausen, Germany

E-mails: \email{e.a.dubovyk@gmail.com}, \email{janusz.gluza@us.edu.pl}, \email{tord.riemann@desy.de},\email{jusovitsch@googlemail.com}
}

\abstract{\hspace*{0cm}Mellin-Barnes (\mb) techniques applied  to  integrals emerging in particle physics  perturbative calculations are summarized. New versions of 
  \ar{} packages  which construct planar and non-planar \mb{} representations are shortly discussed. The numerical package \mbn{} is presented for the first time which is able to calculate with a high precision multidimensional \mb{}
integrals in Minkowskian regions. 
  Examples are given for   massive vertex   integrals which include threshold effects and several scale parameters.   
}

\FullConference{Loops and Legs in Quantum Field Theory - LL 2016, \\
                24 - 29 April 2016            \\
                Leipzig, Germany}

\begin{document}

\section{Introduction}

From a perspective of application of mathematical methods in particle physics, a history of Mellin-Barnes integrals starts with the work
"Om definita integraler" \cite{mellin1895} in which a so-called Mellin transform has been considered 
\begin{equation}
\mathcal{M}[f](s)=\int_0^\infty dx x^{s-1}f(x).
\end{equation}
Here $f(x)$ is a locally integrable function
where $x$ is a positive real number and $s$ is complex in general.

A few years later another paper appeared by Barnes,
"The theory of the gamma function" \cite{barnes1900}. What is nowadays commonly called the Mellin-Barnes representation is a merge of the 
above two: a sum of terms is replaced by an integral representation on a complex plane
 
\begin{eqnarray}
\frac{1}{(A+B)^{\lambda}}&=&
  \frac{1}{\Gamma (\lambda)}
  \frac{1}{2 \pi i}
  \int_{-i \infty}^{+i \infty}dz
  \Gamma (\lambda+z)\Gamma (-z)
  \frac{B^{z}}{A^{\lambda +z}}.
  \label{eq:tool1}
\end{eqnarray}

This relation has an immediate application to physics, for instance, a massive propagator can be written as 

\begin{eqnarray}
\frac{1}{(p^2-m^2)^{a}}&=&
  \frac{1}{\Gamma (a)}
  \frac{1}{2 \pi i}
  \int_{-i \infty}^{+i \infty}dz
  \Gamma (a+z)\Gamma (-z)
  \frac{(-m^2)^{z}}{(p^2)^{a +z}}.
  \label{eq:tool2}
\end{eqnarray}

The upshot of this change is that a mass parameter $m$ merges with a kinematical variable $p^2$ into the ratio $\left(-\frac{m^2}{p^2}\right)^z$, effectively the integral becomes massless.    
Examples how to solve simple Feynman diagrams using this relation can be found in the textbook \cite{Smirnov:2004ym}. In more complicated 
multi-loop cases the introduction of Feynman integrals appears useful
\begin{eqnarray}
\label{eq-scalar1}
G_L[T(k)] &=&
\frac{1}{(i\pi^{d/2})^L} \int \frac{d^dk_1 \ldots d^dk_L~~T(k)}
     {(q_1^2-m_1^2)^{\nu_1} \ldots (q_i^2-m_i^2)^{\nu_j} \ldots
       (q_N^2-m_N^2)^{\nu_N}  }
\\\nonumber
&=& 
\frac{(-1)^{N_{\nu}} \Gamma\left(N_{\nu}-\frac{d}{2}L\right)}
{\prod_{i=1}^{N}\Gamma(\nu_i)}
\int_0^1 \prod_{j=1}^N dx_j ~ x_j^{\nu_j-1}
\delta(1-\sum_{i=1}^N x_i)
\frac{U(x)^{N_{\nu}-d(L+1)/2}}{F(x)^{N_{\nu}-dL/2}}~P_L(T)
.
\end{eqnarray}
In a next step, by generalization of Eq.~(\ref{eq:tool1}) the elements of the Symanzik polynomials $F$ and $U$ are transformed into \mb{} 
representations. 
This procedure has been automatized initially in 
\cite{Gluza:2007rt}.

In Eq.~(\ref{eq:tool1}), applied to Eq.~(\ref{eq-scalar1}),  the Gamma functions play a pivotal role, changing the original singular 
structure of propagators into another one. 

To our knowledge in Quantum Field Theory the Mellin transform has been used for the first time in \cite{Bjorken:1963zz}. Later on,  in the 
seventies of the last century, Mellin-Barnes integrals have been used in the context of asymptotic expansion of Feynman amplitudes in 
\cite{Bergere:1973fq} and Mellin-Barnes contour integrals have been further investigated for finite three-point functions in 
\cite{uss1975}, followed by further related work \cite{Boos:1990rg,Davydychev:1992xr,Usyukina:1993ch}.
However, in terms of mass production of new results in the field, a real breakthrough came by the end of the last millenium when the 
infra-red divergent massless planar two loop box has been solved  analytically 
using \mb{} method \cite{Smirnov:1999gc}, followed in the same year by the non-planar case  
\cite{Tausk:1999vh}.

Presently there are several public software packages for  the application of \mb{} integrals in particle physics calculations. 
On the \mb{} Tools webpage \cite{mbtools} the following codes related to the MB approach can be found:
 \begin{itemize}
\item  The \ar{} project \cite{Gluza:2007rt,Gluza:2010rn,Katowice-CAS:2007} -- for the creation of \mb{} representations. The present \ar{} versions are:
\begin{enumerate}
 \item v1.3 - manual approach, useful for testing
 \item v2.1 - complete, automatic approach for planar diagrams (some Mathematica bugs fixed, improvements concerning factorizations of the 
Symanzik polynomials) -- the loop-by-loop approach (LA method) 
 \item v3.1 -  non-planar diagrams \cite{Blumlein:2014maa,Dubovyk:2015yba} (efficient two-loop constructions and some 3-loop diagrams) -- 
the global approach (GA method)
\end{enumerate}
Appropriate Mathematica examples for \mb{} constructions and improvements can be found in \cite{Katowice-CAS:2007},
fully automatic three-loop version for non-planar cases is under development. 

It is clear that to decide between \la{} and \ga{} methods, knowledge of the planarity of integrals is needed. For this the {\rm PlanarityTest.m} package \cite{Bielas:2013v11,Bielas:2013rja} is used which gives {\rm FALSE} or {\rm TRUE} output concerning planarity of a given diagram.   

\item \mb{} by M. Czakon \cite{Czakon:2005rk} and \verb|MBresolve| by V. Smirnov \cite{Smirnov:2009up} -- for the analytic continuation of 
Mellin-Barnes integrals in $\epsilon$;
\item  \verb|MBasymptotics| by M. Czakon  -- for the parametric expansion of Mellin-Barnes integrals;
\item  \verb|barnesroutines| by D. Kosower   -- for the automatic application of the first and second Barnes lemmas;

\end{itemize}

At the last Loops and Legs conference 
a strategy for possible analytical solutions of \mb{} integrals was outlined using  the \mbsums package which changes \mb{} integrals into 
infinite sums \cite{Blumlein:2014maa}. However, till now there is no real breakthrough in this approach, especially when many-scale 
integrals are concerned. Convergence and summation of an obtained \mb{} sums is intricated \cite{Ochman:2015fho}, even for two-dimensional 
cases \cite{Friot:2011ic}. To use the \mb{} method further on, and applying it to physical processes, if possible in a completely automatic 
way, we have changed the strategy and started to work on an efficient and purely numerical calculation of \mb{} integrals in the 
Minkowskian region. 
 
It is not accidental that in the title the word "region" in plural appears in the context of the Minkowskian kinematics. In various 
kinematic regions specific difficulties emerge in calculation of Feynman integrals due to threshold effects, singularities or several 
mass parameters involved. So, these objects are in general hard in numerical evaluation, though many less (NLO) or more general  approaches 
exist to deal with the problem. They are based on tree-duality, generalized unitarity, reductions at the integrand level, improved 
diagrammatic approach and recursion relations applied to higher-rank tensor integrals, 
simultaneous numerical integration of amplitudes over the phase space and the loop momentum, contour deformations, expansions by regions, 
sector-decomposition. For more general reviews see \cite{Smirnov:2004ym,Anastasiou:2005cb,Freitas:2016sty}. At the one-loop level the 
situation is much simpler and advanced software exists, applied already to many physical processes, such as 
{\rm
  FeynArts/FormCalc}~\cite{Hahn:2000kx,Nejad:2013ina},
{\rm  CutTools}~\cite{Ossola:2007ax}, {\rm
  Blackhat}~\cite{Berger:2008sj}, {\rm
  Helac-1loop}~\cite{vanHameren:2009dr}, {\rm
  NGluon}~\cite{Badger:2010nx}, {\rm Samurai}~\cite{Mastrolia:2010nb},
{\rm Madloop}~\cite{Hirschi:2011pa}, {\rm Golem95C}~\cite{Cullen:2011kv}, {\rm GoSam}~\cite{Cullen:2011ac},
{\rm PJFry}~\cite{Fleischer:2012et}
and {\rm OpenLoops}~\cite{Cascioli:2011va}. Some of them are necessarilly supported by basic one-loop integral libraries \cite{vanOldenborgh:1990yc,vanHameren:2010cp,Ellis:2007qk,Denner:2016kdg}.

Going beyond the one-loop level, so far only few numerical packages are able to deal with Minkowskian regions.
The most advanced programs are based on the sector decomposition approach, 
{\rm Fiesta 4}~\cite{Smirnov:2015mct}, {\rm SecDec 3}~\cite{Borowka:2015mxa}. {\rm NICODEMOS}~\cite{Freitas:2012iu} is based on contour deformations. There are also complete programs 
dedicated specifically to the precise calculation of two-loop self-energy diagrams \cite{Martin:2005qm,Caffo:2008aw}.
These are so far the only public numerical multiloop projects where calculation in Minkowskian regions is feasable, some other proposals have been anounced for instance in  \cite{Ghinculov:1995sd,Ghinculov:1998pd,Becker:2012nk,Becker:2012bi,Passarino:2001wv}. 

\section{Automation in calculations of Mellin-Barnes integrals}

Thanks to the {\rm MB.m} package \cite{Czakon:2005rk}, Mellin-Barnes integrals have been used intensively as numerical cross checks  for 
analytical results obtained in numerous works. Such checks are easily possible in Euclidean space.
The first trial in the direction of numerical integration of \mb{} integrals in Minkowskian space was undertaken in \cite{Freitas:2010nx}. 
The method developed there based on rotations of integration variables in complex planes
has been applied successfully to the calculation of
two-loop diagrams with triangle fermion subloops for the $Z \to bb$  formfactor
 \cite{Freitas:2012sy}. 
 Another approach to numerical integration was considered in \cite{Peng:2012zpa} where the steepest descent method was explored for stationary point contours. It is an interesting direction, though no clear way has been worked out so far for higher dimensional integrals.
  Let us mention that yet another interesting numerical application 
 of \mb{} integrals for phase space integrations can be found in \cite{Somogyi:2011ir} and \cite{Anastasiou:2007qb,Anastasiou:2013srw}. There some parametric integrals are considered and transformations of \mb{} integrals into Dirac delta constraints have been explored. 

In these proceedings a new approach to the numerical calculation of \mb{} integrals is presented which has been developed during the work 
on the  $Z \to bb$ vertex, aiming in evaluation of complete two-loop electroweak corrections to this process, see \cite{tordll2016,Dubovyk:2016aqv}. 
Historically, the {\rm MB.m} package has been developed and first applied in the Bhabha massive QED 2-loop calculations, as a cross-check 
for analytical Master Integrals and their asymptotic expansions \cite{Czakon:2004wm,Czakon:2006pa,Actis:2007fs,Actis:2008br}.  From the 
point of view of \mb{} integrals, the $Z \to bb$ project is more challenging. It gives 3 dimensionless scales in a specific Minkowskian 
region $(s=M_Z^2)$  with a variance of masses
 $M_Z, M_W, m_t, M_H$ involved and  intricate threshold effects.  

There are about ${\cal{O}}(10^3)$ scalar and tensor integrals  to be worked out  for the two-loop electroweak $Z \to bb$ amplitude\footnote{Initial tensors of rank 5 and 4 are reduced easily to objects of maximally rank 3 tensors \cite{tordll2016,Dubovyk:2016aqv}.} so automation is necessary, which goes in two basic steps: 

\begin{enumerate}
\item[1.] Construction of \mb{} representations and analytic continuations;
\item[2.] Numerical integrations.
\end{enumerate}

In the first step, to remind in short the \ar{} project \cite{Gluza:2007rt,Katowice-CAS:2007}, for planar cases the  automatic derivation of \mb{} integrals 
by \ar{} is optimal using the so-called loop-by-loop approach (\la{}).
There, the  simple one-loop $U=1$ in each iterative loop is secured by definition, and concern is on effective $F$ polynomial factorization 
with minimal number of terms. Presently the newest version is \ar{} v2.1 \cite{Katowice-CAS:2007}. 
In the global approach (\ga{}) which is used in non-planar cases, both the $F$ and $U$ 
polynomials are changed into \mb{} representations with help of 
Eq.~(\ref{eq:tool1}) just in one step. On the way a suitable change of Feynman variables is made and the Cheng-Wu theorem is used. 
Presently the newest version is \ar{} v3.1 \cite{Katowice-CAS:2007}.
 
Beyond two-loops, one may choose a hybrid approach which treats planar subloops separately. For such cases  \ar{} v3.1 may be used \cite{Katowice-CAS:2007}. For other 3-loop cases the semi-automatic \ar{} v1.3 may help (user can manipulate itself on optimizing $F$ polynomials manually; without that typically of the order of 20-dimensional \mb{} integrals emerge.

Finally, all the new \ar{} versions have an option to construct MB-integrals in 
dimensions different from $d=4-2 \epsilon$. An example is given in \cite{Katowice-CAS:2007}.

In the second step, a completely new software \mbn{} has been used
\cite{mbnum}.
In the next section some core ideas which made possible to calculate \mb{} integrals in Minkowskian regions used in \mbn{} are given.

\section{Direct numerical integrations of \mb{} integrals in Minkowskian regions.\\ 
}
  
 \subsection{Basic Problems}
  
\mb{} integrals when treated numerically in Minkowskian regions suffer from two kinds of potential problems connected with
  
  \begin{enumerate}
\item[I.] Bad oscillatory behavior of integrands;
\item[II.] Fragile stability for integrations over products and ratios of Gamma ($\Gamma$) functions.
\end{enumerate}

The problem has been discussed initially in \cite{Czakon:2005rk} using a two-loop example factorizing into QED massive vertex integrals

\begin{eqnarray}
V(s)  &=&
\frac{e^{\epsilon\gamma_E}}{i\pi^{d/2}} \int \frac{d^d k}  {[(k+p_1)^2-m^2] [k^2] [(k-p_2)^2-m^2]} ~~=~~ \frac{V_{-1}(s)}{\epsilon} + 
V_0(s) + \cdots
\label{vqed}
\end{eqnarray}

In the above equation the Laurent  series expansion of the integral in $\epsilon$, $d=4-2 \epsilon$ is given. To see the problem, it 
is enough to look at the leading divergent integral  $V_{-1}(s)$, which, translated into the \mb{} representation, takes the following 
form, with $m=1,s=(p_1+p_2)^2$:

\begin{eqnarray}
\nonumber \\ 
V_{-1}(s) &=& 
-
 ~ \frac{1}{2s} 
\int\limits_{-\frac{1}{2}-i \infty}^{-\frac{1}{2}+i \infty} 
\frac{dz}{2\pi i}~~
\underbrace{(-s)^{-z}}_{\bf {Part\; I}}
\overbrace{\frac{\Gamma^3(-z)\Gamma(1+z)} {\Gamma(-2z)}}^{\bf {Part\; II}}
\label{parts}
\end{eqnarray}

Parts I and II refer to basic numerical problems of the general \mb{} integrals connected with kinematical variables and masses of propagators involved. For instance, for diagrams with massless propagators, the Gamma functions in \mb{} integrals include arguments with single variables while in massive cases some of them are multiplied by 2, as in the denominator of Eq.(\ref{parts}).
Fortunately, this massive integral is known in an analytical form for long. Nowaday even computer algebra systems like Mathematica can do 
the job, and summing up residues of Eq.~(\ref{parts}) we get

\begin{eqnarray}
\nonumber \\ 
V_{-1}(s) &=&  
\frac{1}{2}\sum_{n=0}^{\infty}  \frac{s^n} { \binom{2n}{n} (2n+1)}
=
 \frac{2\arcsin(\sqrt{s}/2)}{\sqrt{4-s}\sqrt{s}}.
\label{anal}
\end{eqnarray}

\begin{figure}[h!]
  \centering
  \includegraphics[width=0.75\textwidth]{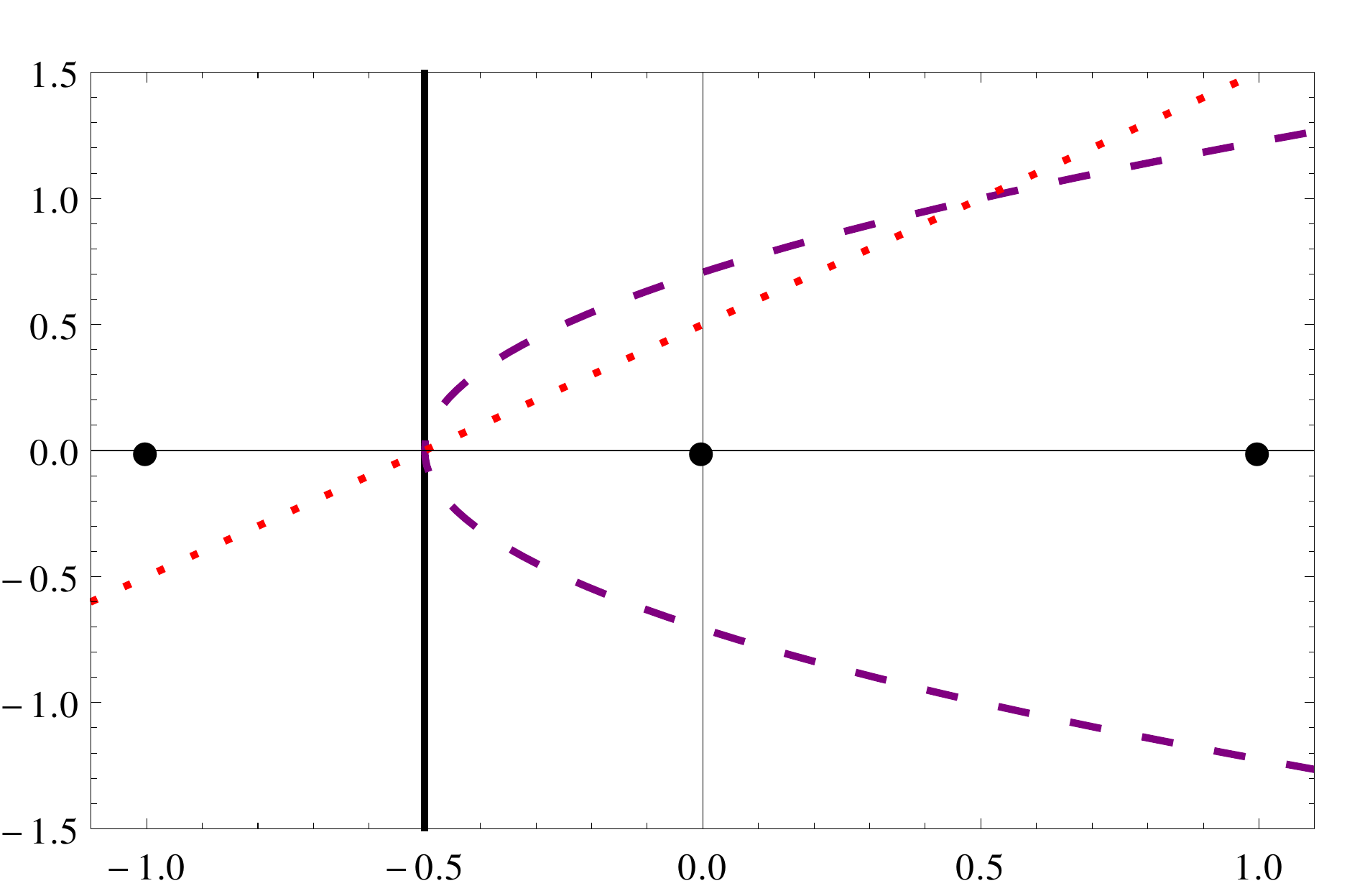}
  \begin{picture}(0,0)
  \put(-220,170){$C_1$}
   \put(-290,65){$C_2$}
   \put(-35,42){$C_3$}
 \put(-350,105){\large $t$}
  \put(-160,210){\large $\Re[z]$} 
   \put(-190,110){\large $\alpha$} 
 \put(-185,107){\oval(35,35)[tr]} 
  \end{picture}
\caption[]{Integration contours chosen for the real part of the complex variable $z$ defined in Eqs.~(\ref{parts}),(\ref{zpar}) and 
Eqs.~(\ref{c1})-(\ref{c3}). For $C_2$ $\alpha = \arctan (\frac{1}{\theta})$. Deformation from $C_1$ to $C_2$ or $C_3$ does not cross poles (black dots).
}
  \label{bild1}
\end{figure}
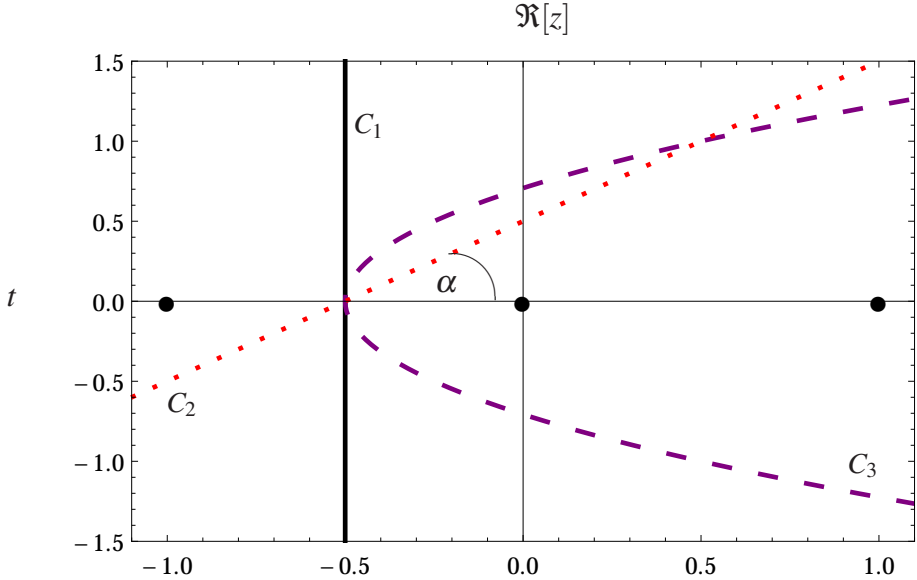

So, we can test numerically some basic ideas connected with contour deformations.
Let us parameterize integral Eq.~(\ref{parts}) as
\begin{equation}
z = \Re [z] + i\; t,~~~t \in (-\infty, +\infty)
\label{zpar}
\end{equation}
\\
where $\Re [z]$ is chosen in three different ways (see Fig.~\ref{bild1}):
\begin{eqnarray}
z(t) = x_0+it~: ~~~~V_{-1}^{C_1}(s) &=& \int_{-\infty}^{+\infty} (i)~dt~ J[z(t)] ,
\label{c1}
\\
z(t) = x_0+ {\bf {\theta}} t+it~: ~~~~ V_{-1}^{C_2}(s) &=& \int_{-\infty}^{+\infty} ({\bf {\theta}}+i)~dt~ J[z(t)] ,
\label{c2}\\
 z(t) = x_0+at^2+it~:~~~~ V_{-1}^{C_3}(s) &=& \int_{-\infty}^{+\infty} (2at+i)~dt~ J[z(t)].\label{c3}
\end{eqnarray}
 
The accuracy of the results of integration at some Minkowskian point depends strongly on the chosen contour.  
For instance, taken $s=2$, Eq.~(\ref{anal}) gives an exact result (which is purely real)

\begin{eqnarray}
V_{-1}(2)|_{\text{analyt.}} &=&  \frac{\pi}{4}  = {0.785398163}39744830962.
\label{analnum}
\end{eqnarray}

If we try to find a numerical solution directly to the integral Eq.~(\ref{parts}) by trying to control oscillatory behavior of the 
integrand using special algorithms, like the Pantis method, as in \cite{Czakon:2005rk}, the obtained result estimated this way is  
$V_{-1}(2)|_{\text{Pantis}}^{MB.m} = 0.7925 
-  {0.0225\; i}$. It is obviously not an acceptable result for further use. 

Let us estimate the result using contours $C_1,C_2,C_3$. We get


\begin{eqnarray}
V_{-1}(2)|_{\text{$C_1$
}} &=& 4.4574554985139977188+  {4.5139812364645122275\;i}
\\
V_{-1}(2)|_{\text{$C_2$}} &=& {0.785398163}3859819 -5.420140575251864\cdot {10^{-15}} \;i
\\
V_{-1}(2)|_{\text{$C_3$}} &=& {0.785398163}2958756 + 2.435551760271437 \cdot {10^{-15}}\;i.   
\end{eqnarray}

As we can see, taking countours $C_2$ and $C_3$ and comparing numerical results with Eq.~(\ref{analnum}), already 10-12 digits of accuracy 
for the integration can be obtained.  
Similar accuracy can be obtained for other  points in the Minkowskian region above the second 
threshold, $s>4$.  

\subsection{Basic methods and tricks for accurate numerical integrations}
 
 We experienced three main methods to integrate efficiently \mb{} integrals, namely

\begin{itemize}
\item[I.] Specific integration methods for oscillating integrands;
\item[II.] Integration contour deformations;
\item[III.] Integration contour shifts.
\end{itemize}

As discussed and shown in the last section, method I is not effective for numerical treatment of \mb{} integrals in physical regions {(and it is known that it is a complicated issue, see for instance \cite{Bailey2012HandtohandCW})} and the method II 
is limited, though it may be quite effective for 1-dimensional cases. 

Method {III}  is new, and as will be shown, it is an effective and programmable method, even for numerical calculation of multi-dimensional 
 \mb{} integrals.  

\begin{center}
{\bf{Method III. The idea.}}
\end{center}

The idea of contour shifts is rather plain and straigthforward. Imagine we have some \mb{} integral with fixed real parts of complex 
integration variables $z_i$ (as it is usually the case, such \mb{} representations are available using \ar{} and {}\mbm). We then shift one 
or more variables $z_i$ by multiply integer numbers. By virtue of Gamma functions and kinematics involved, a new, "shifted" \mb{} integral 
is obtained, plus a bunch of residue integrals from controlled crossing of poles. The aim is to get a shifted original integral whose 
absolute magnitude,  by virtue of applied shifts, is smaller and smaller. How far we go with shifts  (so going down in magnitude of the 
original MB integral)  depends on which accuracy of the final numerical result we aim at. The remaining residue 
\mb{} subintegrals  after shifts are of lower MB dimension. The procedure is iterative. In a next step \mb{} residue integrals of 
lower MB dimensions can be treated the same way. When a procedure is terminated depends   on the accuracy of the   generated residue \mb{} 
integrals and on the desired accuracy for the original \mb{} integral. In passing, there can be large numerical cancellations among 
numerically equal  subintegrals of different sign, which must be also controlled properly.       

As an illustrative example of the efficiency of shifts, let us take the  two-dimensional integrand

\begin{equation}
J(z_{1},z_{2})=\frac{ 2(-\mathrm{\frac{s}{M_{Z}^2}})^{-z_{1} - z_{2}}
  \Gamma[-1 - z_{1} - 2 z_{2}] \Gamma[-z_{1} - z_{2}] \Gamma[-z_{2}] \Gamma[
  1 + z_{2}]^3 \Gamma[1 + z_{1} + z_{2}]}{\mathrm{s^2} \Gamma[1 - z_{1}]}
\end{equation}
\\
and start with the contour of integration $C_1$, Eq.~(\ref{c1}), where $\Re{[z_1]}=z_{10}=0,\;\Re{[z_2]}=z_{20}=-0.7$.
It is interesting to note that at the the kinematic point $s/M_Z^2=1+i\varepsilon$ ($\varepsilon$ is an arbitrarily small parameter 
chosing the correct sheet),
which is a point explored in the  $Z\to b b$ studies \cite{tordll2016,Dubovyk:2016aqv}, shifts works well. To see this,
we shift $z_2$ variable, $z_2=z_{20}+n$. The integral is now a discrete function of the number of shifts ${n}$:
\begin{equation}
I^{C_{1}}(s,M_{Z},{n})=\int\limits_{-\infty}^{+\infty}\int\limits_{-\infty}^{+\infty}(i)^2J(z_{10}+it_{1},z_{20}+{n}+it_{2})\mathrm{d}t_{1}\mathrm{d}t_{2}.
\end{equation}
\\
Using Stirling's formula 
\begin{equation}
\label{eq:gamma2limit}
\Gamma (z) = \sqrt{2\pi}~e^{-z} ~z^{z-1/2}
\left(1 
+ \frac{1}{12 z} + \frac{1}{288 z^2} - \frac{139}{51840 z^3} - \frac{571}{2488320 z^4} 
+ \cdots\right)
\end{equation}
and the relation 
$\ln(-|R|) \to \ln(-|R| \pm i\epsilon) = \ln(|R|) \pm  {i~\pi }$ 
the worst asymptotic behavior for the integrand is for $t_{1}\rightarrow-\infty,\;t_{2}\rightarrow0$:
\begin{equation}
J(z_{10}+it_{1},z_{20}+{n}+it_{2})\simeq t_{1}^{-2-2(z_{20}+n)}.
\end{equation}

\begin{figure}[t]
\centering
  \includegraphics[clip=true,trim=80px 610px 240px 70px,width=.6\textwidth]{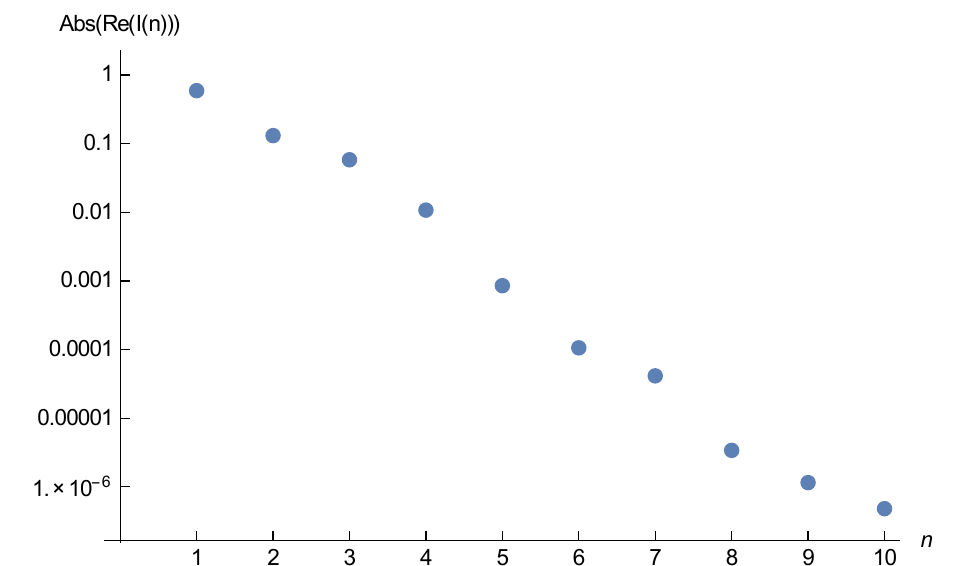}
  \begin{picture}(0,0)
   \put(-205,135){$t_{1}^{-2.6}$}
   \put(-185,120){$t_{1}^{-4.6}$}
   \put(-165,115){$t_{1}^{-6.6}$}
   \put(-145,100){$t_{1}^{-8.6}$}
   \put(-125,80){$t_{1}^{-10.6}$}
   \put(-105,65){$t_{1}^{-12.6}$}
   \put(-85,55){$t_{1}^{-14.6}$}
   \put(-65,35){$t_{1}^{-16.6}$}
   \put(-45,25){$t_{1}^{-18.6}$}
   \put(-25,20){$t_{1}^{-20.6}$}
  \end{picture}
\caption{Module of real part of the integral $I^{C_{1}}(s,M_{Z},{n})$ as a function of $n$.}
\label{johann1}
\end{figure}
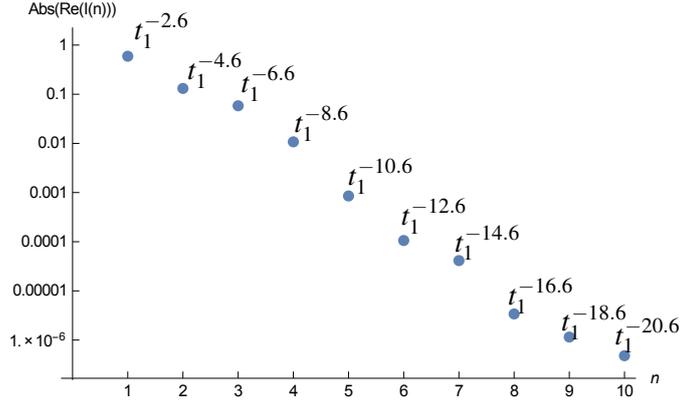

For $n=0$ and $z_{20}=-0.7$, the integrand $J(z_{1},z_{2})$ drops off like $t_{1}^{-0.6}$. This slow convergence is similar to the QED 
massive vertex, discussed above. However, increasing $n$, the module of the integrand becomes smaller, making possible to   get it 
arbitrarily small, see Fig.~\ref{johann1}.
 
We can see that the shifts improve the asymptotic behavior and reduce the order of magnitude of the integral $I^{C_{1}}(s,M_{Z},{n})$.
{{The absolute and the module of an imaginary part of $I^{C_{1}}(s,M_{Z},{n})$ behave similarly.}}

Automatic algorithms for finding the suitable shifts and contour deformations are implemented in MBnumerics.m \cite{mbnum}. 
At the moment an effective strategy is: Starting from original $n-$dimensional \mb{} integrals, MBnumerics.m looks for well 
converging $n-1$ and $n-2$ integrals, and remaining $n-$dimensional integrals. 
Up to 4-dimensional integrals, the deterministic Cuhre method of the CUBA package 
\cite{Hahn:2004fe,Hahn-cuba:2016} can be used. In this way accuracy of calculation can be controlled. 

At the moment it appears that linear contour deformations as in Eq.~(\ref{c2}) are  sufficient (in \cite{Freitas:2016sty,Freitas:2010nx} 
they are called contour rotations) for the evaluation of shifted and residue integrals, when merged with another trick, namely mapping of 
variables in integrands. Contour $C_3$ is the basic contour used for an evaluation of 1-dim integrals. 

Mapping of variables is necessary, making possible the numerical integration of integrals over finite regions. At the same time 
the numerical stability of integrations is improved. In \cite{Czakon:2005rk} a logarithmic mapping has been used

\begin{eqnarray}
z_i &=& x_i + i\ln\left(\frac{t_i}{1 - t_i}\right),~~ t_i \in (0,1),\;\;\;
 {\rm Jacobian}:\;\;
  J_i(t_i) = \frac{1}{t_i(1 - t_i)}.
  \label{eq-jacob}
\end{eqnarray}

\begin{figure}
        \includegraphics[width=.4\linewidth]{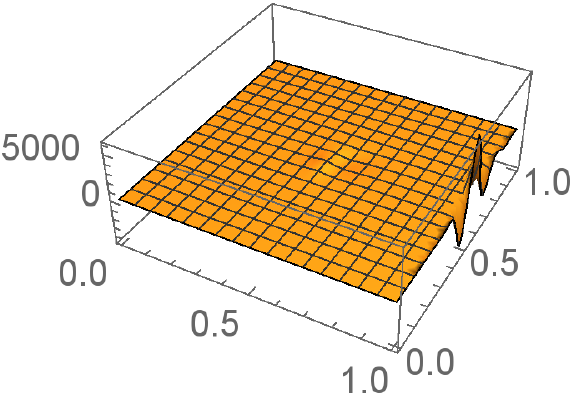}
         \includegraphics[width=.4\linewidth]{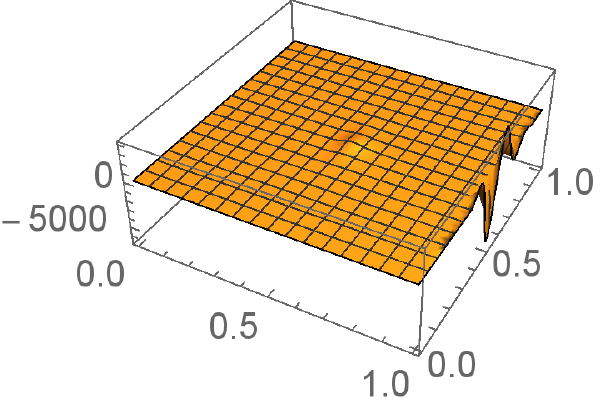}
\caption[]{Logarithmic mapping for the integrand in Eq.~(\ref{intmap}) . On left (right) real (imaginary) part of the integral is given.}
\label{figlog}
\end{figure}

Unfortunately, the curvature rules of Cuhre cannot approximate integrands with a power law {$1/t_i$} behaviour, which is exactly what may happen at the boundaries of the unit hypercube, due to the Jacobians \eqref{eq-jacob}.
An example for such a problematic integral defined in Eq.~(\ref{intmap}) is given in Fig.~\ref{figlog}.

\begin{equation}
I=
\int\limits_{-\frac{1}{3}-i\infty}^{-\frac{1}{3}+i\infty} d z_{1}\int\limits_{-\frac{2}{3}-i\infty}^{-\frac{2}{3}+i\infty} 
dz_{2}
\left(\frac{-s}{M_{Z}^2}\right)^{-z_{1}}
\frac{\Gamma[-z_{1}]^3\Gamma[1 + z_{1}]\Gamma[ z_{1} - z_{2}]\Gamma[-z_{2}]^3\Gamma[1 + 
z_{2}]\Gamma[1 -z_{1} + z_{2}]}
{s~~ \Gamma[1-z_{1}]^2\Gamma[-z_{1} - z_{2}]\Gamma[ 1 + z_{1} - z_{2}]}
\label{intmap}
\end{equation}

Instead of a logarithmic, a tangent mapping is used in \mbn:

\begin{eqnarray*}
z_i&=& x_i  + i~\frac{1}{\tan(-\pi t_i)},~~ t_i \in (0,1),\;\;\;
 {\rm Jacobian}:\;\;
 J_i = \frac{\pi}{\sin^2{[(\pi t_i)]}}.
\end{eqnarray*}

\begin{figure}
        \includegraphics[width=.4\linewidth]{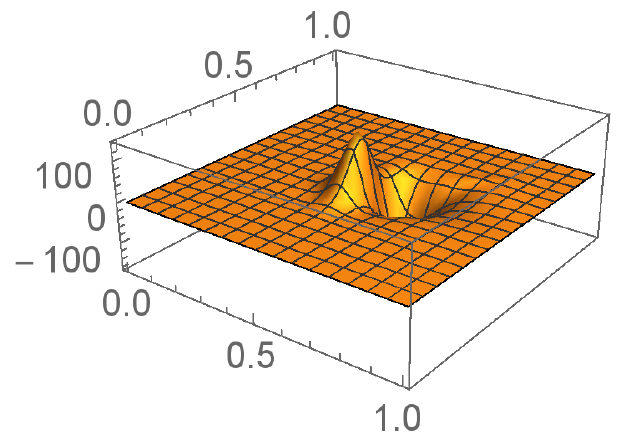}
         \includegraphics[width=.4\linewidth]{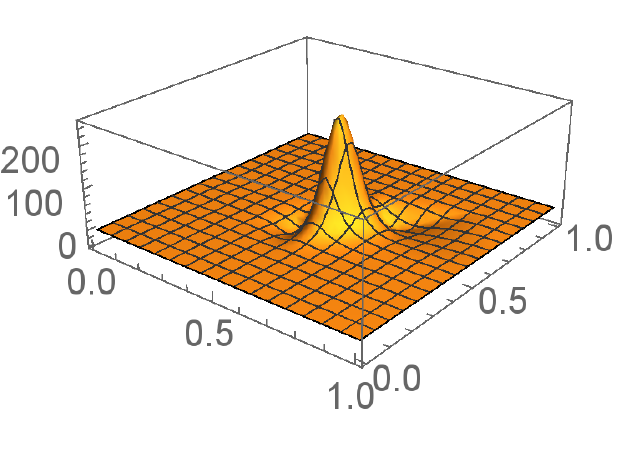}
\caption[]{Tangent mapping for the integrand in Eq.~(\ref{intmap}) . On left (right) real (imaginary) part of the integral is given.}
\label{figtan}
\end{figure}

Comparing Fig.~\ref{figlog} and Fig.~\ref{figtan}, 
even with naked eye one can see that tangent mapping does not give boundary instabilities and the integrand is relatively smooth.
To improve the stability of numerical integrations further, in addition, $\Pi_i \; \Gamma_i \to  e^{\sum_i \ln \Gamma_i}$ transformation helps  
considerable.

As already said, shifts come with mappings and  contour deformations and \mbn{} uses linear contour deformations (rotations), Eq.~(\ref{c2}).
 The point is that a linear change of variables introduces an additional exponential factor  which, chosen properly, may help to damp 
integrand oscillations.
Using this transformation no poles of Gamma functions are crossed as the rotation is applied to all \mb{} integration variables at once, first noted in \cite{Freitas:2010nx}. 
To see how to choose the rotation parameter properly to get damping factors,  let us consider the asymptotic behaviour of the integral 
Eq.~(\ref{parts}) using Eq.~(\ref{eq:gamma2limit}).
\\ 
  { Part I} in Eq.~(\ref{parts}) for the contour $C_1$ gives

\begin{equation}
  (-s)^{-(x_0+i t)} = (s)^{-(x_0+i t)} (-1-i \epsilon)^{-(x_0+i t)}=(s)^{-(x_0+i t)}
 e^{i \pi x_0} {e^{-\pi t}}.
\label{step1}
\end{equation}
We can see that if  $ { t \to - \infty }$ the last exponential factor explodes. 
Fortunately, from Part II of Eq.~(\ref{parts}), we get 

\begin{equation}
\Re \left[ \frac{\Gamma^3[-x_0-i t] \Gamma[x_0+i t]}{\Gamma[2 (x_0+i t)]} \right]  \simeq 2 \pi ^{3/2} \sqrt{\left| \frac{s}{t}\right| }e^{-\pi |t|}. 
\end{equation}
The numerator cancels out an exponential factor in Eq.~(\ref{step1}), though the badly convergent part $t^{-1/2}$ remains. 
It can be stabilized further by rotating the $z$ variable by some angle $\alpha (\theta)$, see Fig.~\ref{bild1}
 \\
\begin{equation}
z= x_0+i t \to x_0+(\theta +i) t.
\end{equation}
 \\
Now, the complete result for Eq.~(\ref{parts}) can be cast in the following way  
 \begin{equation} 
\Re\left[\lim\limits_{|z| \to \infty}V_{-1}(s)\right] \sim
2 \pi ^{3/2} \sqrt[4]{\theta ^2+1} \sqrt{\left| \frac{s}{t}\right| }
 e^{-\pi  \left| t\right| +t \arg (-s)+\theta  t \log
    \left(\frac{4}{s}\right)}
,\;\;s=\Re[s]+i \varepsilon,
\end{equation}
where
\begin{equation} 
\arg(-s) = \begin{cases}
  -\pi, & \Re[s]>0, \\
  -\pi/2, & \Re[s]=0, \\
  0, & \Re[s]<0, 
\end{cases}
\end{equation}

It can be seen easily that for any value of the kinematical variable $s\neq4$ and $s\neq0$ the $\theta$ parameter can be chosen to make {the 
real part of the exponents argument negative}\footnote{For $s=4$, the integral $V_{-1}(s)$ must be considered together with the 
threshold factor $\sqrt{1-\frac{4}{s}}$.}, for instance, {for $s=1+i\varepsilon$ } the condition is  
\begin{equation}
0 < \theta <\frac{\pi }{\log (2)}.
\end{equation}
 \\
Let us look for some more complicated, two-dimensional example of contour deformations:

\begin{equation}
 J(z_{1},z_{2})=\frac{\left(\frac{M_{W}^2}{M_{T}^2}\right)^{z_{2}} s^2 \left(-\frac{s}{M_{T}^2}\right)^
 {z_{1} - z_{2}}
   \Gamma[-z_{1}] \Gamma[z_{1}] \Gamma[2 - z_{2}] \Gamma[ 4 + z_{1} - z_{2}] \Gamma[ z_{2}] \Gamma[-z_{2}])}{4 M_{T}^4  \Gamma[
   6 + z_{1} - 2 z_{2}]}.
   \label{intJ}
\end{equation}
\\
 The integral will be integrated over $C_{1}$:
\begin{equation}
 K^{C_{1}}(s,M_{W},M_{T})=\int\limits_{-\infty}^{+\infty}\int\limits_{-\infty}^{+\infty}(i)^2J(z_{10}+it_{1},z_{20}+it_{2})\mathrm{d}t_{1}\mathrm{d}t_{2},\;\;z_{10}=0.7,\;z_{20}=-1.2 ,
 \end{equation}
 and over $C_{2}$:
 \begin{equation}
 K^{C_{2}}(s,M_{W},M_{T})=\int\limits_{-\infty}^{+\infty}\int\limits_{-\infty}^{+\infty}(i+{\color{blue}\theta})^2J(z_{10}+(i+{\color{blue}\theta})t_{1},z_{20}+(i+{\color{blue}\theta})t_{2})\mathrm{d}t_{1}\mathrm{d}t_{2},\;\;z_{10}=0.7,\;z_{20}=-1.2.
 \end{equation}
 \\
The worst asymptotic behavior of the integrand $J^{C_{1}}$ takes place for $t_{1}\rightarrow 0,\;t_{2}\rightarrow-\infty$
 \begin{equation}
 J(z_{10}+t_{1},z_{20}+t_{2})\simeq t_{2}^{-\frac{3}{2}}.
 \end{equation}
 
\begin{figure}[t]
 \centering
   \includegraphics[width=.4\linewidth]{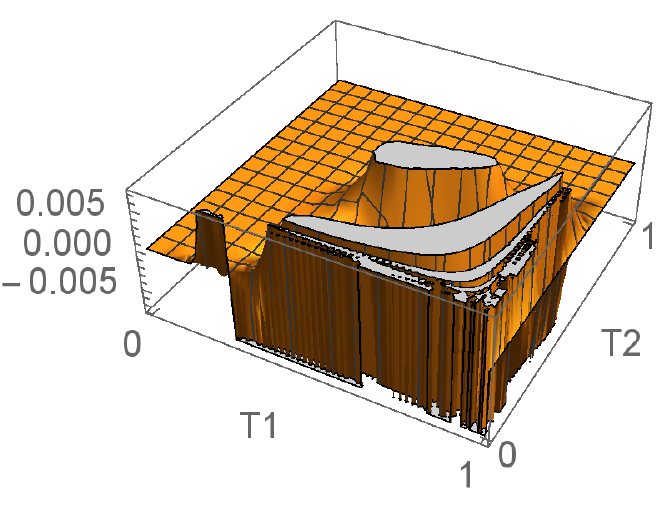}
   \includegraphics[width=.4\linewidth]{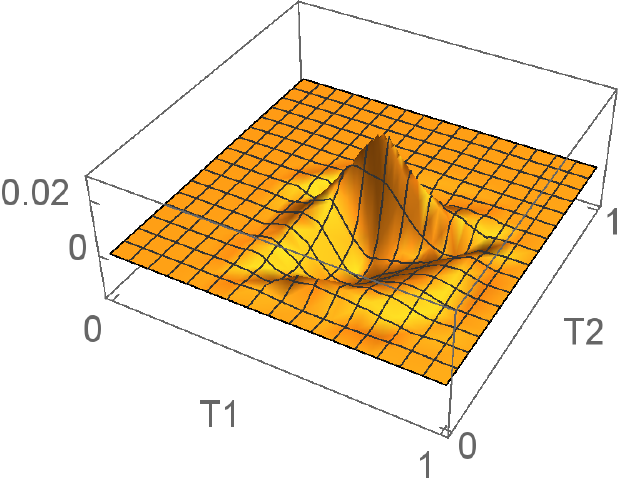}
   \label{johann2}
   \caption{Real part of the integrand $J$ defined in Eq.(3.22)
   evaluated over contours $C_1$ and $C_2$. $T_{1}$ and $T_{2}$ are integration variables due to a tangent mapping: $t_{i}=1/\tan[-\pi 
T_{i}]$.}
\label{figrparts} 
 \end{figure}

Thus the integral $K^{C_{1}}$ is very slowly convergent.
Taking $s=M_Z^2$ and $\theta=0.7$, the asymptotic behavior of the integrand $J^{C_{2}}$ is like in the Euclidean case and the numerical 
evaluation of $K^{C_{2}}$ yields honest high accuracy. In Fig.~\ref{figrparts} real parts of the integrand $J$ evaluated over contours $C_1$ 
and $C_2$ are  given.

\subsection{\mbn{}, present situation}  

The algorithm has been applied succesfully so far to up to four-dimensional \mb{} integrals with the desired accuracy (8 digits).
In calculations, the  \ar{} constructions are not the main issue as far as time of calculation is concerned. For numerical results, time 
consuming is the determination of optimal contours where the proper grid for threshold kinematics and the treatment of tails of integrands 
bother. The second time factor is connected with numerical integration over the optimal contours. To get high accuracy,
the optimal strategy is to treat \mb{} integrals with  maximal four dimensions, and 
to use Cuhre, which is not a Monte Carlo, but a deterministic algorithm \cite{Hahn-cuba:2016}. A decent precision can be obtained in this 
way. {For some specific cases \mb{} integrals have been initially reduced using KIRA package \cite{kira}, followed by their numerical evaluation with \mbn.} 

\begin{figure}
\begin{center}
\includegraphics[scale = 1]{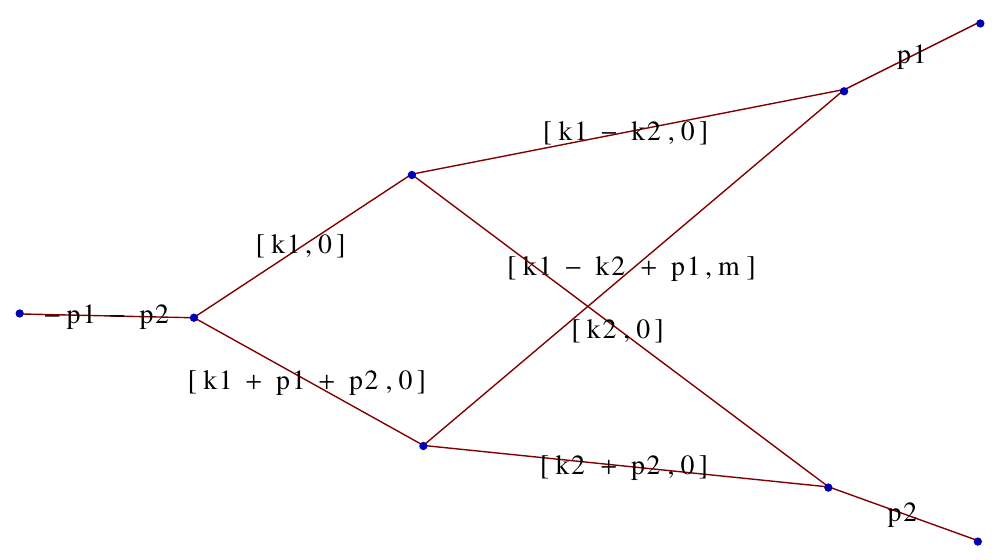}
\end{center}
\caption{Non-planar vertex with one massive crossed line. Figure generated by \pltest{} \cite{Bielas:2013rja,Bielas:2013v11}.}
\label{fignp}
\end{figure}

Finally, we give as another example the constant part of  the 3-dimensional integrand Eq.~(\ref{exint}) drawn in Fig.~\ref{fignp} 

\begin{eqnarray}
&& (-s)^{-2 {\epsilon}-{z_2}-2} \left(m^2\right)^{{z_2}}  
\Gamma [-{\epsilon}]  \Gamma
   [-{z_1}] \Gamma [-{z_2}]  \Gamma [-{z_3}] \Gamma^2[{z_3}+1] \Gamma[-{\epsilon}-{z_1}] \Gamma[-{\epsilon}-{z_2}] \Gamma[{z_1}+{z_3}+1] \nonumber \\
      &&\nonumber \\
 & & \times  \frac{ \Gamma[-2{\epsilon}-{z_1}-{z_3}-1] \Gamma[-2
   {\epsilon}-{z_2}-{z_3}-1] \Gamma[-2{\epsilon}-{z_1}-{z_2}-{z_3}-1] 
   \Gamma[2 {\epsilon}+{z_1}+{z_2}+{z_3}+2]}
   {\Gamma[-2 {\epsilon}-{z_1}] \Gamma[-3 {\epsilon}-{z_2}] 
   \Gamma[-2 {\epsilon}-{z_2}] \Gamma[-2
   {\epsilon}-{z_1}-{z_2}]},
   \nonumber   \\&&
   \label{exint}
\end{eqnarray}
\\
which shows how  powerful \mbn{} can be.
In this case, results obtained with different available methods and programs in the Euclidean region  are the following, 
$-(p_1+p_2)^2=m^2=1$:
{
$$
\begin{array}{ll}
{\rm Analytical}: &       -0.4966198306057021 \\ 
{\rm MB(Vegas)}:   &      -0.4969417442183914 \\     
{\rm MB(Cuhre)}:    &     -0.4966198313219404 \\
{\rm FIESTA}:     &       -0.4966184488196595  \\ 
{\rm SecDec}:    &        -0.4966192150541896 
\end{array} 
$$
}
\\
For the Minkowskian region, $(p_1+p_2)^2=m^2=1+i\varepsilon$, constant part:
$$
\begin{array}{lc}
{\rm Analytical}:  &    -{0.778599608}979684 - {4.123512}593396311 \cdot i \\
{\rm MBnumerics}:  &    -{0.778599608}324769 - {4.123512}600516016 \cdot i\\
{\rm MB(Vegas)}:   &                         {\rm big\; error} \\
{\rm MB(Cuhre)}:   &                          {\rm  NaN} \\
{\rm FIESTA}:      &                         {\rm big\; error}\\
{\rm SecDec}:      &                          {\rm big\; error}
\end{array}
$$

 For more examples and comparisons, see \cite{tordll2016,Dubovyk:2016-tobepublished}.
 
\section{\label{secsummary}Summary}

We have summarized the present status of the \ar{} project for the construction of \mb{} integrals. New versions of the package for planar 
and non-planar integrals are given.

The new package \mbn{} has been discussed in which \mb{} integrals can be evaluated numerically in a Minkowskian region. Oscillatory 
behaviour of \mb{} integrals is treated by shifts of variables, stabilility of integrals is further improved by contour deformations and 
mapping of variables. Shifts of contours are so powerful that sometimes the method alone is sufficient to obtain high accuracy.  Difficult 
cases like thresholds are also treatable now. For  $Z \to b b$, \mbn{} turned out to be a very strong and effective tool 
\cite{tordll2016,Dubovyk:2016aqv}.     
\section*{Acknowledgements}
The work of I.D. is supported by a research grant of Deutscher Akademischer Austauschdienst (DAAD) and
by Deutsches Elektronensychrotron DESY. The work of J.G. is supported by the Polish National Science
Centre (NCN) under the Grant Agreement No. DEC-2013/11/B/ST2/04023. The work of T.R. is supported
in part by an Alexander von Humboldt Polish Honorary Research Fellowship. The work of J.U. is supported
by Graduiertenkolleg 1504 "Masse, Spektrum, Symmetrie" of Deutsche Forschungsgemeinschaft (DFG). We would like to thank Peter Uwer and his Group ``Phenomenology of Elementary Particle Physics beyond the Standard Model'' at Humboldt-Universit\"{a}t for providing computer resources.
 
\providecommand{\href}[2]{#2}
\addcontentsline{toc}{section}{References}


\end{document}